\documentstyle[12pt]{article}

\begingroup\makeatletter\ifx\SetFigFont\undefined
\def\x#1#2#3#4#5#6#7\relax{\def\x{#1#2#3#4#5#6}}%
\expandafter\x\fmtname xxxxxx\relax \def\y{splain}%
\ifx\x\y   
\gdef\SetFigFont#1#2#3{%
  \ifnum #1<17\tiny\else \ifnum #1<20\small\else
  \ifnum #1<24\normalsize\else \ifnum #1<29\large\else
  \ifnum #1<34\Large\else \ifnum #1<41\LARGE\else
     \huge\fi\fi\fi\fi\fi\fi
  \csname #3\endcsname}%
\else
\gdef\SetFigFont#1#2#3{\begingroup
  \count@#1\relax \ifnum 25<\count@\count@25\fi
  \def\x{\endgroup\@setsize\SetFigFont{#2pt}}%
  \expandafter\x
    \csname \romannumeral\the\count@ pt\expandafter\endcsname
    \csname @\romannumeral\the\count@ pt\endcsname
  \csname #3\endcsname}%
\fi
\fi\endgroup

\advance \topmargin by -\headheight
\advance \topmargin by -\headsep

\evensidemargin \oddsidemargin
\begin{document}

\centerline{\large\bf Thermal Effective Potential of the O(N) 
Linear $\sigma$ Model}
\vskip 2 cm
\begin{center}
{\bf Giovanni AMELINO-CAMELIA}\\
\end{center}
\begin{center}
{\it Theoretical Physics, University of Oxford,
1 Keble Rd., Oxford OX1 3NP, UK}
\end{center}
\vskip 1 cm
\centerline{\bf ABSTRACT }
\medskip
The finite-temperature effective
potential of the $O(N)$ linear $\sigma$ model is studied,
with emphasis on the implications for the investigation of
hot hadron dynamics.
The contributions from all the ``bubble diagrams''
are fully taken into account for arbitrary $N$;
this also allows to address
some long-standing issues concerning 
the use of non-perturbative approaches
in (finite-temperature) field theory.


\vfill
\noindent{OUTP-97-10P \space\space\space hep-ph/9702403
\hfill February 1997}

\newpage
\pagenumbering{arabic}
\setcounter{page}{1}
\pagestyle{plain}
\baselineskip 12pt plus 0.2pt minus 0.2pt


The temperature-induced phase transition
of the $O(N)$ Linear $\sigma$ Model (L$\sigma$M)
has recently attracted renewed interest,
especially as a model of the chiral phase transition
that might be triggered 
by certain ultrarelativistic collisions of hadrons or nuclei.
While some of the relevant experimental signatures,
such as the ones associated
to {\it disoriented chiral condensates}
(see, {\it e.g.}, Refs.~[1-3]),
are best estimated within the framework of non-equilibrium
(classical or quantum) field
theory~[3-8],
there is also considerable interest 
(see, {\it e.g.}, the recent Refs.~[9-11])
in the quasi-equilibrium/thermal description
of the chiral phase transition.
Moreover, some of the scenarios of phenomenological interest
consist~\cite{BdVH,LANL} 
of a first stage in which thermal equilibrium is reached,
followed by a stage in which the system
evolves out of equilibrium.
The present note is devoted 
to the study of the ``thermal" (finite-temperature)
effective potential of the $O(N)$ L$\sigma$M,
which is an essential theoretical tool
for the analysis of quasi-equilibrium/thermal properties.

The difficulties involved in the evaluation of 
thermal effective potentials
have been confronting the physics community for decades.
They were first encountered in the study of certain condensed-matter
problems, but even for what concerns those problems
a general consensus on the procedures to be followed
has not yet been reached
(see the recent Ref.~\cite{okumura} and references therein).
In the context of the relativistic quantum field theories
of interest to particle physics
the same, essentially infrared, problems associated to
the description of thermal effects
are accompanied by the familiar ``pathology" of ultraviolet divergencies.
The need to resort to non-perturbative approaches
in the analysis of thermal effective potentials
was already emphasized to the particle-physics community
more than 20 years ago, in the works~\cite{doja}
that provided technical support for the
Kirzhnits-Linde~\cite{lindePT}
proposal of ``symmetry restoration'' (thermal field fluctuations at
sufficiently high temperatures restore
symmetries broken {\it a la} Higgs in vacuum).
In particular,
it was immediately clear that at high temperatures
the ``bubble diagrams"~\cite{doja,kapunpb,kapu} (see, Fig.~1)
of any order in the coupling constants
could not be neglected in any consistent approximation,
even if the coupling constants are small,
as a result of the fact that ``bubble subgraphs"
contribute a factor proportional to the square of the temperature.
This has motivated the development of various approaches
to the resummation of such bubble diagrams.
[In the case of the $O(N)$ L$\sigma$M
as a model of hadron dynamics the coupling constant 
is itself large, and 
even at small temperatures it is necessary
to study the non-perturbative sector.]

In the study of relativistic field theories with $N$ scalar fields,
the proposed approaches to bubble-resummation can be divided in two
broad categories; attempts aiming at
the exact resummation of all bubble diagrams,
sometimes referred to as Hartree approaches,
and attempts aiming at
the resummation of only those bubble diagrams
that would {\it survive} the $N \rightarrow \infty$ limit,
usually referred to as large-$N$ approaches\footnote{The
conventional large-$N$ approach for theories in
vacuum (at zero temperature) also involves some non-bubble (``sunset-type")
contributions; however,
at high temperatures, in light of
the above-mentioned fact that the bubble diagrams are dominant,
it makes sense to limit the large-$N$ approach to bubble diagrams.
It is this truncated version that is qualified as large-$N$ approach
in the present note.}.
After 20 years of debate only rather recently
some consensus has emerged on the Hartree side;
in fact, the independent studies~\cite{hsugac,quigac}
have compared various Hartree approaches\footnote{The 
studies \cite{hsugac,quigac} focused on diagrammatic techniques useful 
in the analysis of scalar theories.~A technique based on
the analysis
of ``hard thermal loops''~\cite{htl} can also be very useful 
in the study of thermal effective potentials, but its natural
framework is the one of gauge theories.~Moreover, 
it is well known that the Hartree approach
can be implemented in a non-diagrammatic way, through
the use of gaussian wave functionals~\cite{gaussian}.}
proposed for the study of thermal effective potentials,
and found that only 
the one proposed in Ref.\cite{pigac}
by Pi and this author (which was based on the CJT
formalism~\cite{cjt})
and 
the one proposed in Ref.~\cite{lindeannph} 
by Kirzhnits and Linde
lead to reliable results.
Much less controversy has affected large-$N$ analyses
of the thermal effective potential 
(see, {\it e.g.}, Refs.~\cite{doja,vlargen});
in fact, the large-$N$ limit significantly simplifies the counting
of multiloop bubble diagrams, which is the most common source of
inaccuracy.

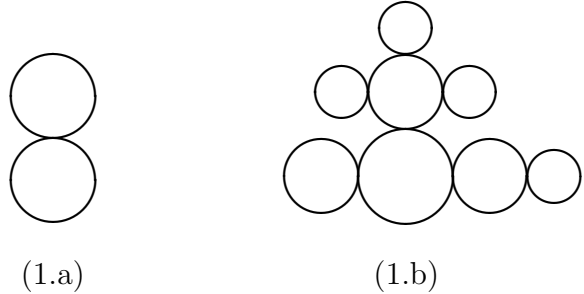
\begin{figure}
\begin{picture}(7218,2000)(2993,-3500)
\thicklines
\put(5400,-2125){\circle{500}}
\put(5400,-2650){\circle{500}}
\put(5200,-3300){\makebox(0,0)[lb]{\smash{\SetFigFont{12}{14.4}{rm}(1.a)}}}
\put(7600,-1700){\circle{300}}
\put(8000,-2100){\circle{300}}
\put(7200,-2100){\circle{300}}
\put(7600,-2100){\circle{450}}
\put(8525,-2625){\circle{300}}
\put(8125,-2625){\circle{450}}
\put(7075,-2625){\circle{450}}
\put(7600,-2625){\circle{600}}
\put(7400,-3300){\makebox(0,0)[lb]{\smash{\SetFigFont{12}{14.4}{rm}(1.b)}}}
\end{picture}
\caption{As examples of ``bubble diagrams",
the ``2-bubble diagram" is drawn in (1.a)
and one of the ``8-bubble diagrams" is drawn in (1.b).}
\end{figure}

In modeling (part of) the hadron dynamics 
associated to ultrarelativistic collisions of hadrons or nuclei
the $O(4)$ L$\sigma$M appears to be most relevant~\cite{krishna},
and the large-$N$ approach might be too drastic when $N \! = \! 4$.
Unfortunately,
the historically poor
understanding (reviewed in the recent Ref.~\cite{matsui})
of the ultraviolet structures 
encountered when performing the complete  
resummation of bubble diagrams with $N \! > \! 1$
has been obstructing the way of analyses within the Hartree thermal
effective potential.
It is perhaps because of these problems that even recent
studies of the $O(4)$ L$\sigma$M at finite temperature
have chosen to dispose of ultraviolet divergences in
one or another {\it ad hoc} (unjustified)
fashion (see, {\it e.g.} Ref.~\cite{matsui})
or drop quantum effects altogether (see, {\it e.g.} Ref.~\cite{jorgen}).
However, the ``ultraviolet problems" of the Hartree approach
are to be ascribed to poor handling of the difficult formalism,
rather than being inherent; in fact,
this author's recent investigation of
the ultraviolet structure of the Hartree (thermal) effective potential
in a theory with two scalar fields~\cite{gacjt2f} has shown
that this potential is indeed renormalizable when
the contributions from all the bubble diagrams are
included (with correct counting and symmetry factors)
and the interdependence of the gap equations for the
dressed propagators is properly taken into account.

In the present note the formalism developed in \cite{pigac,cjt,gacjt2f}
is used to obtain the Hartree bubble-resummed thermal
effective potential of the $O(N)$ L$\sigma$M
for arbitrary $N$,
and in particular the ultraviolet sector is found to
be renormalizable.
Besides providing a useful tool for future investigations
of the L$\sigma$M at finite temperature,
this result also allows to investigate some of the differences
between the Hartree and the large-$N$ approach
(of course, the exact resummation of all bubble diagrams
performed in the following for arbitrary $N$ also entails,
upon appropriate limiting procedure, the corresponding
large-$N$ approximation).
In particular, it is emphasized below that,
although both approaches are renormalizable,
when considered as effective low-energy descriptions, with finite cut-off,
they manifest a rather different dependence on the cut-off.
Also prominent among the issues debated in relation to the differences
between Hartree and large-$N$ is Goldstone's theorem,
which is generally expected 
to be verified in the large-$N$ approach,
but violated in the Hartree approach.
However,
this expectation emerged from analyses of unrenormalized Hartree equations
(see, {\it e.g.} \cite{BdVH,LANL}), and
it is emphasized below that renormalization
can modify substantially the structures relevant for
Goldstone's theorem.

Having set the agenda for the present note, let us start 
its computational part by writing explicitly 
the Lagrangian density of the $O(N)$ L$\sigma$M
\begin{equation}
L = - {1 \over 2} (\partial_{\mu} \sigma) (\partial^{\mu} \sigma) -
{1 \over 2} (\partial_{\mu} \vec{\pi}) (\partial^{\mu} \vec{\pi}) 
- {m^2 \over 2} \left( \sigma^2 + \vec{\pi}^2 \right)
- {\lambda \over 6 N}  \left( \sigma^2 + \vec{\pi}^2 \right)^2
~,
\label{lfzz}
\end{equation}
where the pion vector field is understood to have $N \! - \! 1$ components.

As clarified in Refs.~\cite{matsui,hsugac,quigac,pigac,gacjt2f,pap8},
the most convenient method of evaluation of
the thermal effective potential including the contribution from
the complete (Hartree) sum of bubble diagrams emerges within 
the CJT formalism\cite{cjt,consolicjt,jackbanf}
and involves the bubble-resummed (dressed) 
propagator $D_\bigcirc (\phi;k)$:
\begin{eqnarray}
V_\bigcirc (\phi) =
W_\bigcirc [\phi;D_\bigcirc (\phi;k)] \equiv
V_{tree}
- {1 \over 2} \, \hbox{$\sum$}\!\!\!\!\!\!\!\int_k \, \ln D_\bigcirc
+ {1 \over 2} \, \hbox{$\sum$}\!\!\!\!\!\!\!\int_k \, 
[D_{tree}^{-1} D_\bigcirc -1] 
+ W^*[\phi;D_\bigcirc ]
~,\label{hfbubble}
\end{eqnarray}
where $\phi$, the ``classical field'', is the usual c-number 
argument of the effective potential,
$V_{tree}$ and $D_{tree}$ are the tree-level potential and
propagator respectively, and
$W^*$ includes all the ``2-bubble diagrams'' ({\it i.e.}
diagrams with the topology of two rings touching at one point
as in Fig.~1.a) with lines representing $D_\bigcirc (\phi;k)$.
The bubble-resummed 
propagator $D_\bigcirc (\phi;k)$
can be obtained~\cite{cjt} as the function that stationarizes
$W_\bigcirc [\phi;D]$ with respect to $D$-variations.
Also notice that
the conventional {\it sum-integral} notation
is used in (\ref{hfbubble}) to denote
the combination of integration over
spatial components of the momentum
and sum over Matsubara frequencies characteristic of 
finite-temperature field theory in the imaginary time formalism.

In the case of the $O(N)$ L$\sigma$M
one is interested in
the effective potential $V$ as a function
of the classical field associated to the
shift $\{ \sigma, \vec{\pi} \} \rightarrow \{ \sigma 
+ N^{1/2} \phi, \vec{\pi} \}$;
in fact, in physical applications the potential is tilted in the $\sigma$
direction by a chiral symmetry-breaking term, which is ignored
here for simplicity. (The factor $N^{1/2}$ is convenient,
since it renders~\cite{cjp} 
the symmetry-breaking value of $\phi$
essentially independent of $N$.)
The Hartree bubble-resummed
effective potential
corresponding to this $\sigma$ shift
can be written as
\begin{eqnarray}
V_\bigcirc &=& {N {m}^2 \over 2} \phi^2 +
{N {\lambda} \over 6} \phi^4
+{1 \over 2} \, 
\hbox{$\sum$}\!\!\!\!\!\!\!\int_k \, 
\{ \ln [k^2+M_\sigma^2] + (N-1) \ln [k^2+M_\pi^2] \}
\nonumber\\
& & - {1 \over 2} \,  
[M_\sigma^2 \! - \! m^2 \! 
- \! 2 \lambda \phi^2] \, P[M_\sigma]
- {N - 1 \over 2} \,  
[M_\pi^2 \! - \! m^2 \! 
- \! {2 \lambda \over 3} \phi^2] \, 
P[M_\pi]
\nonumber\\
& &
+ {\lambda  \over 2 N} \left( P[M_\sigma] \right)^2
+ {\lambda \over 6}  {N^2 -1 \over N} \left( P[M_\pi] \right)^2
+ {\lambda \over 3} {N - 1 \over N} P[M_\pi] P[M_\sigma]
~,
\label{vsumazzt}
\end{eqnarray}
where for convenience $V_\bigcirc$
has been expressed (without loss of generality)
in terms of the dressed masses $M_\sigma$ and $M_\pi$
rather then the dressed propagator $D_\bigcirc$.
The relation between $M_\sigma$, $M_\pi$
and $D_\bigcirc$
is $[D_\bigcirc^{-1}]_{\Psi,\chi} \equiv 
\delta_{\Psi \sigma} \delta_{\chi \sigma} (k^2 + M_\sigma) +
\delta_{\Psi \pi} \delta_{\chi \pi} (k^2 + M_\pi)$,
and from the above-mentioned stationarization requirement
for the dressed propagators one obtains in the case
of the $O(N)$ L$\sigma$M the following ``gap equations"
for the dressed masses\footnote{While stopping short of investigating
the Hartree thermal effective potential,
the recent work \cite{matsui} did report candidate gap equations
for the dressed masses in the case $N \! = \! 4$.~Those gap equations
are however 
incorrect, as the reader can easily see by comparison
with (\ref{gapthzzt}).~As in other instances in the related literature,
the inaccuracies in the gap equations of Ref.~\cite{matsui} can 
be associated to the handling of the pion degrees of freedom 
as if they were not independent.~Of course they are independent, but 
ultimately the 
symmetries provide the pions with a common dressed mass $M_\pi$.}
\begin{eqnarray}
M_\sigma^2 &=& m^2 + 2 \lambda \phi^2 
+ {2 \lambda \over N}  P[M_\sigma] 
+ {2 \lambda \over 3} {N-1 \over N} P[M_\pi] 
~,
\label{gapthzzt}\\
M_\pi^2 &=& m^2 
+ {2 \lambda \over 3} \phi^2 
+ {2 \lambda \over 3} {N+1 \over N} P[M_\pi] 
+ {2 \lambda \over 3 N} P[M_\sigma]
~.
\label{gapthzztb}
\end{eqnarray}
Here and in (\ref{vsumazzt}), $P[M]$ denotes
the ``tadpole"~\cite{doja,kapu}
\begin{eqnarray}
P[M] \!\! & \equiv & \!\! \hbox{$\sum$}\!\!\!\!\!\!\!\int_p  
\, [p^2 + M^2(\phi;p)]^{-1} = I_1 - M^2 I_2 + P_f[M]
\label{gxxb}\\
P_f[M] \!\! & \equiv & \!\! {M^2 \over 16 \pi^2} 
\ln {M^2 \over \mu^2} 
- \int {d^3k \over (2 \pi)^3}~
\left[\sqrt{|{\bf k}|^2 + M^2}  
\left( 1 - exp \left( { \sqrt{|{\bf k}|^2 
+ M^2} \over T} \right) \right) \right]^{-1}
~,
\label{gxxa}
\end{eqnarray}
where $\mu$ is an arbitrary
(renormalization) scale,
and $I_{1}$ and $I_{2}$ have been separated out as contributions
that diverge in the limit of infinite ultraviolet
momentum cut-off $\Lambda$
\begin{equation}
I_1 = \Lambda^2/ (8 \pi^2) ~,~~~I_2 = 
(\ln \Lambda^2 / \mu^2)/(16 \pi^2)
~.
\label{renml}
\end{equation}

The conventional argument concerning violations of Goldstone's theorem
in Hartree is reflected within this formalism in the fact that $M_\pi$
does not vanish at a symmetry-breaking ($\phi \! \ne \! 0$) minimum of the 
potential, as {\it formally} encoded in the (unrenormalized) relation
\begin{equation}
{d V_\bigcirc \over d \phi} = {\partial V_\bigcirc \over \partial \phi} +
{\partial V_\bigcirc \over \partial M} 
{\partial M \over \partial \phi} =
{\partial V_\bigcirc \over \partial \phi} \propto \phi \left[ M_\pi^2 +
{4 \over 3} {\lambda \over N} (P[M_\sigma] - P[M_\pi]) \right]
~.
\label{golda}
\end{equation}
Instead, to the same level of analysis, one finds that Goldstone's theorem
does hold in the large-$N$ approach, 
as implied by the $N \! \rightarrow \! \infty$
limit of Eq.~(\ref{golda}).

In proceeding toward the renormalization of
the effective potential and gap equations
it is important to realize that 
the divergences originate from the ``tadpole'' discussed above
and the ``one loop''
\begin{eqnarray}
\hbox{$\sum$}\!\!\!\!\!\!\!\int_k \, 
\ln [k^2+M^2] \!\! & = & \!\! 
- {M^4 \over 4}  I_2
+{M^2 \over 2}  I_1 + Q_f[M]
\label{voneTreg}\\
Q_f[M] \!\! & \equiv & \!\! {M^4 \over 64 \pi^2}             
[\ln {M^2 \over \mu^2} - {1 \over 2}] 
+ T \int {d^3k \over (2 \pi)^3}~
\ln \left[ 1- 
exp \left( { \sqrt{|{\bf k}|^2 + M^2} \over T} \right) \right]
~.
\label{voneTregb}
\end{eqnarray}
Renormalized ({\it i.e.} finite in
the $\Lambda \! \rightarrow \! \infty$ limit)
expressions
are obtained upon introducing renormalized coupling
and mass that are related to the bare ones by 
\begin{eqnarray}
{m_R^2 \over \lambda_R} = {m^2 \over \lambda} +
{2 \over 3} {N \! + \! 2 \over N} I_1 ~,~~~
{1 \over \lambda_R} = {1 \over \lambda} +
{2 \over 3} {N \! + \! 2 \over N} I_2
~.
\label{renormcoupl}
\end{eqnarray}
It is important that,
given a positive value of the renormalized coupling $\lambda_R$,
the bare coupling is positive only for $\Lambda$ smaller
than the $\Lambda_{L}$ corresponding to the
Landau pole\footnote{This behavior
is related to the well-known ``triviality" of the
theory~\cite{triviality}, {\it i.e.}
$\lambda_R \! \rightarrow \! 0$
as $\Lambda \! \rightarrow \! \infty$
if one insisted on $\lambda \! > \! 0$.}.
For $\Lambda \! > \! \Lambda_{L}$ the bare coupling
is negative, and in particular $\lambda \! \rightarrow \! 0^-$
as $\Lambda \! \rightarrow \! \infty$.
The renormalizability of the Hartree bubble-resummed
finite-temperature effective potential and gap equations
is manifest in the fact that,
with some tedious algebra here spared to the reader,
one can rewrite the Eqs.~(\ref{vsumazzt})-(\ref{gapthzztb})
as sums of 
terms that are proportional to the bare coupling
({\it i.e.} terms that vanish in the $\Lambda \! \rightarrow \! \infty$ limit)
and terms that involve only renormalized quantities:
\begin{eqnarray}
M_\sigma^2 &=& m_R^2 + {2 \lambda_R \over 3} {N \! + \! 2 \over N^2}
\left ( N \phi^2 + P_f[M_\sigma] \right)
+ {2 \lambda_R \over 3} {N^2 \! + \! N \! - \! 2 \over N^2} P_f[M_\pi] 
\nonumber\\
& & + 
{2 \lambda \over \lambda_R} {N \! - \! 1 \over N^2} (M_\pi^2 - M_\sigma^2)
+ {4 \lambda \over 3} {N^2 \! + \! N \! - \! 2 \over N^3}
\left ( N \phi^2 + P_f[M_\sigma] - P_f[M_\pi] \right)
~,
\label{gaprensigma}\\
M_\pi^2 &=& m_R^2 + {2 \lambda_R \over 3} {N \! + \! 2 \over N^2}
\left ( N \phi^2 + P_f[M_\sigma] \right)
+ {2 \lambda_R \over 3} {N^2 \! + \! N \! - \! 2 \over N^2} P_f[M_\pi] 
\nonumber\\
& & + 
{2 \lambda \over \lambda_R} {1 \over N^2} (M_\sigma^2 - M_\pi^2)
- {4 \lambda \over 3} {N \! + \! 2 \over N^3}
\left( N \phi^2 + P_f[M_\sigma] - P_f[M_\pi] \right)
~,
\label{gaprenpi}
\end{eqnarray}
and (ignoring an irrelevant $\phi$-independent contribution)
\begin{eqnarray}
V_\bigcirc &=& Q_f[M_\sigma] + (N \! - \! 1) Q_f[M_\pi]
+ {N \over 2} \phi^2 M_\sigma^2 
+ {3 \over 4} {N \over N \! + \! 2} {m_R^2 \over \lambda_R}
[M_\sigma^2 + (N \! - \! 1) M_\pi^2]
\nonumber\\
& & 
- {3 \over 8} {N \over 
N \! + \! 2} {M_\sigma^4 + (N \! - \! 1) M_\pi^4
\over \lambda_R} 
+ {1 \over 4} {N \! - \! 1 \over N} (M_\pi^2 - M_\sigma^2)
\left( N \phi^2 + P_f[M_\sigma] - P_f[M_\pi] \right)
\nonumber\\
& & 
+ {3 \over 8} {N \! - \! 1 \over N \! + \! 2} 
{(M_\sigma^2 - M_\pi^2)^2 \over \lambda_R} 
- {N {\lambda} \over 3} \phi^4
~.
\label{vren}
\end{eqnarray}
If one was exclusively interested in a proof of the
renormalizability of the Hartree bubble-resummed 
thermal\footnote{It is worth bringing to the attention 
of those readers who are unfamiliar with the related literature 
the fact that renormalizability in a non-perturbative approach
to a thermal field theory is rather non-trivial.
The theorems/proofs we usually advocate to discuss renormalizability
are inherently perturbative, and bare no implications for 
non-perturbative approaches. Moreover, when dealing with a 
thermal field theory one can encounter (and they have been
encountered in this note) ultraviolet-divergent contributions that
depend on the temperature, and it is non-trivial to find that
there is a temperature-independent (as required by
consistency~\cite{kapu,gacjt2f,jackbanf}) renormalization scheme,
such as (\ref{renormcoupl}).} 
effective potential the terms proportional
to the bare coupling $\lambda$ in the
Eqs.~(\ref{gaprensigma})-(\ref{vren})
could have been simply 
dropped since they vanish in the
infinite cut-off limit.
However, in 
physical analyses of the $O(N)$ L$\sigma$M,
besides the renormalizability of the approach,
which provides an indication of overall consistency
and is also important at the formal level,
one is also interested in the structure of the theory
with a finite cut-off.
In fact, this is motivated mathematically by the implications
of the above-mentioned Landau-pole structure (in 
the $\Lambda \! \rightarrow \! \infty$ limit
the theory is unstable),
and phenomenologically by the fact that the $O(N)$ L$\sigma$M 
is useful only as a low-energy effective description of
hadron dynamics.
With a finite cut-off (small enough to keep at a safe distance 
from the Landau pole~\cite{BdVH,LANL})
all the terms proportional
to $\lambda$ in the Eqs.~(\ref{gaprensigma})-(\ref{vren})
are to be taken into account and lead to the unwanted result
that the Hartree thermal
effective potential is rather sensitive to the value 
of the cut-off.
It is worth emphasizing that,
as manifest in the Eqs.~(\ref{gaprensigma})-(\ref{vren}),
all the explicit dependence on the cut-off 
drops out in the large-$N$ limit.

Let us now go back to the issues concerning Goldstone's theorem.
It is important to notice that after renormalization Eq.~(\ref{golda})
takes the form
\begin{equation}
{d V_\bigcirc \over d \phi} \propto \phi \left[ M_\pi^2 
+ {8 \lambda \over 3} {\phi^2 \over N} 
+ {4 \lambda \over 3} {N \! + \! 2 \over N^2} 
(P_f[M_\sigma] - P_f[M_\pi]) 
+ {\lambda \over \lambda_R}
{2 \over N} (M_\pi^2 - M_\sigma^2)
\right]
~.
\label{goldb}
\end{equation}
Clearly, renormalization has changed things quite dramatically:
in the $\Lambda \! \rightarrow \! \infty$ 
($\lambda \! \rightarrow \! 0^-$)
limit one finds from (\ref{goldb}) that $M_\pi$
vanishes at a symmetry-breaking ($\phi \! \ne \! 0$) minimum of the 
potential, as required by Goldstone's theorem.
However, as clarified above, the behavior of the ``renormalized''
($\Lambda \! \rightarrow \! \infty$) theory is not very relevant
to phenomenological applications, and
Eq.~(\ref{goldb}) 
does reflect a violation of Goldstone's theorem
when the cut-off is finite.
Therefore the generic expectation expressed in the literature,
which relies on the analysis of unrenormalized relations,
appears to hold in this weaker sense after renormalization is performed.
Again, the large-$N$ approach is better behaved,
since the $N \! \rightarrow \! \infty$
limit of Eq.~(\ref{goldb}) is in agreement with
Goldstone's theorem.
It is this author's opinion that
the analysis here reported suggests
that the fundamental difference between
the Hartree and the large-$N$ approaches
resides in the above-mentioned different sensitivity
on a finite cut-off, rather than the widely publicized
issues related to Goldstone's theorem.
In fact, the different properties Hartree and large-$N$ have 
in relation to Goldstone's theorem
are all encoded in the cut-off-dependent
contributions to Eq.~(\ref{goldb}).

It is tempting to read into (\ref{vren}),
and particularly its property (\ref{goldb}),
also something about the different behavior
with temperature of 
the Hartree and the large-$N$ approaches.
As mentioned in the opening of this note,
the bubble diagrams are dominant,
providing a justification for the Hartree approach,
only at high temperatures.
However, if $N$ is indeed large, than the
(large-$N$-type) bubble diagrams are the dominant ones independently
of the temperature.
One would therefore expect the reliability of the large-$N$ approach
to be rather insensitive on the temperature, whereas
the Hartree approach should get better as the temperature increases.
While the computations presented in the present note do not go far enough
(and this author's insight is not deep enough)
to confirm these expectations, it is worth observing that
the deviation from Goldstone's theorem encoded in
Eq.~(\ref{goldb}) becomes less substantial as the temperature increases
(in particular, $\phi^2$ and $P_f[M_\sigma] \! - \! P_f[M_\pi]$
decrease as the temperature increases).
In any case, the renormalized Hartree bubble-resummed thermal
effective potential derived in the present note for arbitrary $N$
should be useful in future investigations of 
the $O(N)$ L$\sigma$M only when the temperatures
of interest ({\it e.g.}, in setting up studies of type~\cite{BdVH,LANL})
are not too close to the critical temperature; in fact,
neither Hartree nor large-$N$ can be trusted
for temperatures very close to the critical temperature,
where any
type of resummation of bubble diagrams becomes insufficient~\cite{pap8}
({\it e.g.}, if these inaccuracies are not taken into account
it would appear that Hartree predicts a first order phase transition
for some theories known to undergo a second order phase transition).
The choice concerning whether to use
the full Hartree bubble-resummed potential
or simply its large-$N$ truncation depends very much
on the problem to be investigated, particularly the
relevant value of $N$
and the role played by the Goldstone modes.
However, the result reported in the present note
can be useful also in those instances
in which the large-$N$ is preferred, since it provides a rough estimate
of the effects being neglected by the leading order in $1/N$.
This is particularly important in light of the fact that
in most cases one is unable to estimate subleading orders
in $1/N$, and the intuition developed in the conventional small-coupling
perturbative approaches cannot be trusted
to apply to the $1/N$ expansion.

\vglue 0.6cm
\leftline{\Large {\bf Acknowledgements}}
\vglue 0.4cm
It is a pleasure to acknowledge conversations
on these and related issues
with J.D.~Bjorken,
D.~Boyanovsky,
F.~Cooper,
R.~Jackiw, 
S.~Larsson,
O.~Philipsen,
and 
S.-Y. Pi.
This work was supported in part by PPARC.

\baselineskip 12pt plus .5pt minus .5pt

\end{document}